\documentclass[preprint2, usenatbib]{emulateapj}
\usepackage{graphics}
\usepackage{epsfig}
\usepackage{natbib}

\def\be{\begin{equation}}
\def\ee{\end{equation}}
\def\ba{\begin{eqnarray}}
\def\ea{\end{eqnarray}}

\shorttitle{High speed satellites in MOND}
\shortauthors{Llinares et al.}

\begin{document}


\title{Physics of Galactic Colliders: high speed satellites in $\Lambda$CDM vs MONDian cosmology}
\author{Claudio Llinares}
\affil{Astrophysikalisches Institut Potsdam, An der Sternwarte 16, 14482
  Potsdam, Germany}
\email{cllinares@aip.de}

\author{HongSheng Zhao}
\affil{SUPA, University of St Andrews, KY16 9SS, UK\footnote{hz4@st-andrews.ac.uk, presently on sabbatical in Leiden
Observatory}}
\affil{Universit\'e de Strasbourg, CNRS, Observatoire astronomique, F-67000 Strasbourg, France}
\email{hz4@st-andrews.ac.uk}

\author{Alexander Knebe}
\affil{Astrophysikalisches Institut Potsdam, An der Sternwarte 16, 14482
  Potsdam, Germany}
\affil{Universidad Aut\'onoma de Madrid, Cantoblanco, 28039 Madrid, Spain}
\email{aknebe@aip.de}

\begin{abstract}
  The statistics of high speed satellite galaxies, as reported in
  the recent literature, can be a powerful diagnosis of the
  depth of the potential well of the host halo, and hence discriminate between competing gravitational theories.  Naively one expects that high speed satellites
  are more common in Modified Newtonian Dynamics (MOND) than in cold
  dark matter (CDM) since an isolated MONDian system has an
  infinite potential well, while CDM halos have finite potential wells. In this \textit{Letter} we report on an initial test of this
  hypothesis in the context of the first generation of cosmological
  simulations utilizing a rigorous MONDian Poisson solver. We find that such high speed encounters are approximately a
  factor of four more common in MOND than in the concordance
  $\Lambda$CDM model of cosmic structure formation.
\end{abstract}


\section{Introduction}
The standard $\Lambda$CDM model \citep[cf.,][]{2008arXiv0803.0547K}
explains the formation of cosmological structure in the non-linear regime in a
hierarchical way, i.e. large structures are not formed monolithically
but by the successive merging of smaller structures
\citep[e.g,.][]{1985ApJ...292..371D}. Recent cosmological simulations
also support the idea of hierarchical formation in MOND gravity 
(\cite{2008arXiv0809.2899L}, but see also
analytical models of \citealt{{2008MNRAS.386.1588S}, {2008ApJ...686.1019Z}}).
The hierarchical merging scenario naturally promotes the picture that
we should observe collisions of galaxies. The question that immediately
arises is what is the nature of the distribution of the relative speed of such
encounters.  Observationally there is evidence that some of these
collisions actually occur with speeds that are not readily reproduced by
simulations of $\Lambda$CDM structure formation \citep{Hayashi06,
  2007MNRAS.380..911S, 2008MNRAS.386.1029K}. There is, for example,
the famous ``Bullet cluster'', an extremely high velocity merger
between two galaxy clusters whose relative speed is between 2500 and 4500 km/sec
depending on the interpretation of the shock speed and the method used to
infer the collision speed (observations/models, analytical/numerical,
N-body/hydro simulations, e.g. \citealt[][]{{2008MNRAS.384..343N},
  {2007MNRAS.380..911S}, {2006ESASP.604..723M}, {2007arXiv0704.0094Z}}).  At
first sight, the upper limit of this interval is too high and may be a problem
for the standard $Lambda$CDM model, but \cite{Hayashi06} showed using the Millenium cosmological simulation \citep{2005Natur.435..629S} that the probability of such an event albeit low, is not zero.

There are a number of such high speed encounters in the literature.  One example of such a collision is the so-called ``line-of-sight Bullet'', i.e. Abel
576, with a relative velocity of 3300~km/sec
\citep{2007ApJ...668..781D}. Furthermore the
``Cosmic Train Wreck'' Abel 520 is a collision with a velocity of
approximately 1000~km/sec \citep{2007ApJ...668..806M}.  The ``Dark
Matter Ring'' cluster Cl0024+17 has a speculated impact velocity of
3000~km/sec \citep{2007ApJ...661..728J} and MACS J0025.4-1222
has two merging components whose relative velocity was measured to
be 2000~km/sec \citep{2008ApJ...687..959B}.   In comparison the random dispersion of velocities in these clusters is only about 500-1000~km/sec.  

A consequence of any high speed collision of mass concentrations seems to
be the decoupling or offsetting of the baryons from the dark
component.  Assuming this being the case, additional examples of collisions
are given in \citet{2005ApJ...618...46J, 2005ApJ...634..813J} but see also 
\cite{2008MNRAS.385.1431H}.  
This kind of objects, with offsets between baryon and DM components, have become what
could be considered as yet another important standard test that any theory for
gravity should pass before being seriously considered \citep[e.g.][]{Will93}.  Simply applying the MOND formula to a universe populated
only with baryons seems to fail this test.  Possible solutions could come from many on-going efforts to
embed the MOND idea in a relativistic framework by adding
complementary (vector) fields besides the standard Einstein's metric
\cite[][]{{2004PhRvD..70h3509B}, {2007PhRvD..75d4017Z},
{2007ApJ...671L...1Z}, {2008arXiv0811.3465Z}} or by
the addition of neutrinos of various kind \cite[][]{{2008MNRAS.383..417A}, {2008A&A...480..313F}, {2008arXiv0811.3465Z}}.  
However we must have in mind, that the same data on the Bullet Cluster would have rejected general relativity
in its original formulation without introducing one or more dark
matter components plus a cosmological constant.


The question that arises from all these data is how to match the low probability
of high-speed encounters predicted by \cite{Hayashi06} for the $\Lambda$CDM model with the
fact that this type of collisions {\it seems} to be common in the observable
universe.  A clue comes from the MONDian point of view where the situation seems to be more
favorable for high velocities.   Previous authors have noted the deep potential in MOND 
 \citep[][]{{2008MNRAS.383..417A}, {2007arXiv0704.0094Z}, {2008MNRAS.384..343N}} is helpful in the context of the Bullet cluster.  On
a smaller scale, high-velocity stars have been studied in the context
of the escape speed in the Milky Way \citep{2008arXiv0809.2087P}.  It was found that MOND can
retain stars of higher velocity than CDM, and the MONDian escape
velocity is more consistent with the RAVE data in the solar
neighbourhood \citep[][]{{2007MNRAS.377L..79F},{2007ApJ...665L.101W}}.  Further, the revised (yet
still discussed) speed of the Magellanic Clouds also favors MOND
\citep[][]{2008MNRAS.386.2199W}.  The
question that previous authors cannot address is how to obtain 
a self-consistent strength of the external field in MOND since they lack
a full cosmological simulation. And as MOND is a non-linear theory it
violates the strong equivalence principle and hence it is {\it mandatory to
simulate galaxies within the cosmological framework and not in
isolation}.

The aim of this paper is not to go further in an explanation of this kind of
systems using MONDian ideas, but to study the consequences of a MONDian
cosmological toy model 
on the probability of such high speed encounters.  
In order to do this, we study the velocity distribution of
substructure extracted from cosmological simulations that have been run using standard and modified gravity.  We show that high speed collisions are more
frequent in MOND than in the concordance $\Lambda$CDM model.

\section{Simulations}\label{simulations}
The analysis presented in this \textit{Letter} is based upon a set of
two simulations published in \citet{2008arXiv0809.2899L}, i.e. the
$\Lambda$CDM and the OCBMond2 model, respectively. Both simulations were run in a  box with a side length
of $32h^{-1}$Mpc, using $128^3$ particles.  They were both run with a
modification of the $N$-body code \texttt{MLAPM}
\citep{2001MNRAS.325..845K}. The $\Lambda$CDM model employs a
background cosmology parametrized by $\Omega_{dm+b}=0.3$,
$\Omega_{\Lambda}=0.7$, and a normalisation of the power spectrum of
the density perturbation of $\sigma_8=0.88$. For the MONDian simulation,
we chose an open universe with neither dark matter nor dark energy but characterized by
$\Omega_{b}=0.04$.  In order to arrive at a
comparable evolutionary stage to the $\Lambda$CDM model at redshift
$z=0$ we had to lower the normalisation $\sigma_8$ to $0.4$ due to the
faster growth of structures in MOND \citep[cf.][]{2001ApJ...560....1S,
  2004MNRAS.347.1055K, 2008arXiv0809.2899L}. Both simulations were
started at redshift $z=50$ and used a Hubble constant $H_0=70$ km/sec/Mpc.

We used the MPI version of the \texttt{AHF} halo
finder\footnote{\texttt{AHF} is freely available from
  \texttt{http://www.aip.de/People/aknebe}}
\texttt{AMIGA}'s-Halo-Finder \citep{Knollmann09} to identify
objects, which is based on the \texttt{MHF} halo finder of
\citet{2004MNRAS.351..399G}. For the identification of substructure we
employed the tool \texttt{MergerTree} that comes with the \texttt{AHF}
software package. This algorithm was originally designed to follow halos through
time by tracking the membership of individual particles, but it can be also
used to locate the subhalos of a given host. Since particles that belong to subhalos will belong also to the corresponding host, constructing a merger tree of a halo catalogue with itself will
provide us with a ``subhalo tree'' (rather than a merger tree).  It is
important at this moment to make a remark about our terminology.  We use the term subhalo to refer to the largest substructures embedded in host haloes.  The
mass ratio between our host halos and the most massive subhalo have a median
of 0.23 and 0.15 for the MONDian and Newtonian simulations respectively.  These
numbers are in the range of typical mass ratios for collisions in mergers
of host halos and are well above the typical ratio between hosts and real
substructures (e.g. \citealt{2008ApJ...679.1260M}).  In order to not contaminate our result with unvirialized
objects we further prune our halo catalogue by removing objects with a
high virial ratio ending up with 64 and 58 objects in the Newtonian and
MONDian simulations, respectively. 

For more details regarding these simulations, we refer
the reader to \citet{2008arXiv0809.2899L}

\section{Analysis}
While the primary focus of this \textit{Letter} is the distribution of
the relative velocity of two colliding systems, we still need to
define a proper normalisation for these velocities to correct for the
fact that a more massive host system will lead to a larger
acceleration towards its centre. While others referred to the
rotational velocity at the virial radius of the host for this purpose
\citep[e.g.,][]{Hayashi06}, we rather use the mass-averaged velocity dispersion.

\subsection{Velocity dispersion - Mass relation for MOND and CDM}

In the Newtonian case, the velocity dispersion scales with the mass $M$ as follows:  
 $\sigma \propto V_{cir} =\sqrt{GM/R} \propto R \propto M^{1/3} \propto M_{baryon}^{1/3}$, where we used $M \propto \bar{\rho} R^3$ where $\bar{\rho}$ is the background density, which depends only on redshift.  A similar scaling relation between velocity
dispersion $\sigma$ and mass $M$ can be easily obtained for deep
MOND $\sigma \propto V_{cir} \propto (GM_{baryon} a_0)^{1/4} \propto M_{baryon}^{1/4}$ for a spherical isolated body.  Although not rigorous, we find that this scaling holds fairly well as a mass-averaged total dispersion of the system even in the intermediate MOND regime. 

\begin{figure}
\plotone{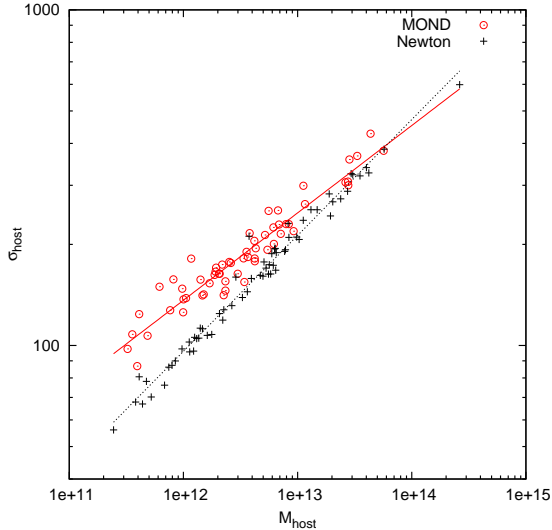}
\caption{Correlations of velocity dispersion ($\sigma_{host}$ in km/s) vs. effective host mass ($M_{host}= 7.25 M_{host, baryon}$ in solar masses) for virialized objects of various sizes in MONDian (circles) 
and Newtonian simulations (pluses).   The lines are fits with arbitrary
normalization for theoretically predicted scaling relations $\sigma_{host}
\propto M_{host,baryon}^{1/4}$ in MOND and $\sigma_{host} \propto
M_{host,baryon}^{1/3}$ in Newtonian; the best fit slopes and very close to 1/4
and 1/3 respectively.  Note that the value 7.25 scales the MOND
baryoninc mass to the Newtonian baryon plus halo mass.}
\label{fig:sigma_m}
\end{figure}

Figure~\ref{fig:sigma_m} shows the $\sigma-M$ relation for the host
systems selected in both our simulations. The lines indicate power laws
fits, whose index agrees closely with the theoretical values 1/4 and 1/3 for MONDian and Newtonian theory, respectively. The fitted normalisation is higher than the theoretical one, owning to
the fact that the simulated halos break the hypothesis of constant density used
in the theoretical approach.

\subsection{Normalising the relative velocities}\label{normalisation}
Special care must be taken when comparing
Newtonian dark matter simulations to collisionless MONDian
simulations, especially when it comes to ``haloes''. While we set out
to use the velocity dispersion of the host system as the normalisation
of the collision velocity in order to account for the mass of the
host, we have just seen that $\sigma-M$ relations scales differently
in MONDian than in Newtonian physics.  What we need to do now is to move
both simulations into the same theoretical framework (e.g., CDM) by applying
the same technique outlined in \citet{2008arXiv0809.2899L}, i.e. we
divide the MONDian (host) masses by the baryon fraction (${\Omega_{dm+b} \over
  \Omega_b} = {0.3 \over 0.04} = 7.25$) to mock dark
matter haloes for direct comparison to the $\Lambda$CDM model.  From 
these ``MONDian dark matter halo plus baryon masses''  we apply the Newtonian $\sigma-M$
relation to obtain the appropriate $\sigma$ value to compute the 
normalized $V_{rel}/\sigma$.  Note that we actually could reverse this procedure and
transform the Newtonian dark matter haloes into the MONDian frame and
use the MONDian $\sigma-M$ relation to obtain the normalized velocity.  We have applied both methods to our data and the
results are consistent with each other. Hence we show the results only for
the former method to facilitate direct comparison to other CDM simulations.  

\section{Results}
\begin{figure}
\plotone{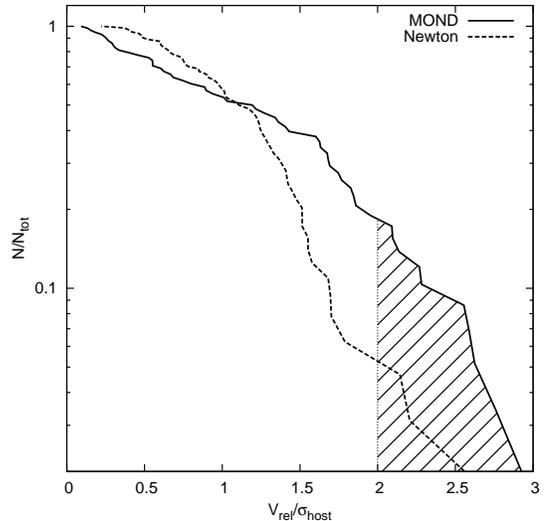}
\caption{Accumulated probability of the relative velocity $V_{rel}$ between our host halos and their most massive subhalo normalized with the effective dispersion $\sigma_{host} = 0.01 (7.25 M_{host,baryon})^{1/3}$ km/sec   (see text for explanation).}
\label{fig:histo}
\end{figure}

Having identified each subhalo and its corresponding host
(cf. Section~\ref{simulations}), we calculate the cumulative
probability distribution of the relative velocity between host and its
most massive subhalo $V_{rel}$. The result is shown in
figure~\ref{fig:histo}.  We can see that the two theories have strikingly different behaviours.  Both allow high speed satellites but there is a stronger tail towards
high-speed in the MONDian case.  The shaded lines show the high
speed region, where the MONDian probability is about four times the Newtonian.

The normalisation of the relative speed was chosen in order to
eliminate the increased acceleration towards more massive host
systems. The credibility of the approach detailed in
Section~\ref{normalisation} can now be verified by simply dividing our
sample into different mass bins. We confirm that this does not lead to
different results even though we decided to not show them in this
\textit{Letter}; we basically recover the same curves as seen in
figure~\ref{fig:histo}.

Further, our results appear to be robust against slight changes in
redshift, i.e.  we neither observe a change in the fact that MONDian
velocities are bigger nor are our results contaminated
by the fact that we may capture collisions at a particular time of
accidentally high velocity. The latter is confirmed by analysing the
simulations at $z=0.036$ leading to an indistinguishable plot. The same conclusion
is reached when we experiment with other plausible normalizations or compute
the distributions of un-normalised relative speed.

As a final test we compare our results against a Newtonian model that does not contain a cosmological constant $\Lambda$, i.e. the open OCBM model of \citet{2004MNRAS.347.1055K} characterized by $\Omega_m=0.04$ and $\Omega_{\Lambda}=0.0$. We acknowledge (though not explicitly shown here) that the relative velocity distribution of the OCBM model is akin to the $\Lambda$CDM model presented in figure~\ref{fig:histo}; we therefore ascribe the differences found in that plot to the effects of MOND rather than the (missing) cosmological constant.

\section{Conclusions}
Inspired by the observational evidence for high-speed encounters of
galaxy clusters we studied the velocity distribution of collisions present
in two cosmological simulations, a standard $\Lambda$CDM model as
well as MOND.  While there may be a problem for $\Lambda$CDM to
accomodate such extraordinary events \citep[e.g.,][]{Hayashi06} we set out
to quantify the probabilty for them in MOND. Within the
limitations of our simulations, we find that there are substantial
differences in the collision velocity of objects in the standard model
of cosmology and its (possible) MONDian counterpart. We observe a much
greater likelihood for high-speed collisions in MOND and therefore
argue that this statistic can be used as a discriminator for the two
competing theories.

We further verify numerically the velocity
dispersion-mass relation for deep MOND gravity whose slope is different to the Newtonian case ($\sigma\propto M^{1/4}$ for
MOND instead of $\sigma\propto M^{1/3}$ for Newtonian physics).
There is a mild scatter about these relations.

We close with a cautionary note: the box size of our
simulation ($32h^{-1}$Mpc) is too small to find objects
\textit{directly} comparable to systems like the Bullet cluster. The
collisional velocity expected for the Bullet cluster ($V_{rel}/\sigma\approx 2.04$) is close to the limit of what we resolve
in figure~\ref{fig:histo}.  While the current result is interesting, and the rescaled $V_{rel}/\sigma$ is likely
insensitive to details of the simulation setup, more simulations (e.g., with a
possible neutrino component and a cosmological constant in MOND) are required to understand how our prediction depends on the cosmological model employed.

Nevertheless, our results here may have far-reaching implications as
well.  Historically, Dark Matter and MOND are competing theories.
Recent studies argue that MOND is a prescription for interactions of a
coupled Dark Energy-Dark Matter field.  Effectively MOND is made by a
non-uniform Dark Energy field.  The places where this field condenses are identified as dark halos.  Our results here could argue that we might
differentiate between theories with interacting Dark Matter-Dark Energy
vs. classical $\Lambda$CDM using data of high speed encounters.  It is encouraging
that some subtle differences on how the Dark Sector self-interacts could leave signatures on ``large astronomical colliders" (LAC).

\section*{Acknowledgements}
This work was carried out under the HPC-EUROPA++ project (project
number: 211437), with the support of the European Community - Research
Infrastructure Action of the FP7 ``Coordination and support action''
Programme.  CL and AK acknowledge funding by the DFG under grant KN
755/2. AK further acknowledges funding through the Emmy Noether
programme of the DFG (KN 755/1).  CL thanks hospitality at Strasbourg
Observatory and HZ to the AIP.  We acknowledge Noam Libeskind for a carefull reading of the manuscript.


\end{document}